\documentclass{article}

\usepackage{PRIMEarxiv}

\usepackage[utf8]{inputenc} 
\usepackage[T1]{fontenc}    
\usepackage{hyperref}       
\usepackage{url}            
\usepackage{booktabs}       
\usepackage{amsfonts}       
\usepackage{nicefrac}       
\usepackage{microtype}      
\usepackage{lipsum}
\usepackage{fancyhdr}       
\usepackage{graphicx}       
\graphicspath{{media/}}     

\title{A Summary of Privacy-Preserving Data Publishing in the Local Setting
}

\author{
 Wenjun~Lin, Jiahao~Qian, Wenwen~Liu, Lang~Wu \\
  Department of Computer Science \\
  National Taiwan University \\
}

\begin{document}
\maketitle

\begin{abstract}

The exponential growth of collected, processed, and shared microdata has given rise to concerns about individuals’ privacy. Consequently, various laws and regulations have been established to oversee how organizations handle and safeguard microdata. One such method is Statistical Disclosure Control, which aims to minimize the risk of exposing confidential information by de-identifying it. This de-identification is achieved through specific privacy-preserving techniques. However, a trade-off exists: de-identified data can often lead to a loss of information, which might impact the accuracy of data analysis and the predictive capability of models. The overarching goal remains to safeguard individual privacy while preserving the data's interpretability, meaning its overall usefulness. Despite advances in Statistical Disclosure Control, the field continues to evolve, with no definitive solution that strikes an optimal balance between privacy and utility. This survey delves into the intricate processes of de-identification. We outline the current privacy-preserving techniques employed in microdata de-identification, delve into privacy measures tailored for various disclosure scenarios, and assess metrics for information loss and predictive performance. Herein, we tackle the primary challenges posed by privacy constraints, overview predominant strategies to mitigate these challenges, categorize privacy-preserving techniques, offer a theoretical assessment of current comparative research, and highlight numerous unresolved issues in the domain.
\end{abstract}


\section{Introduction}
The right to privacy has been a significant discussion topic in computing since the late 60s. This period saw the development of legal safeguards tailored for the digital age, reflecting the longstanding concern over individual privacy. With the current data-intensive environment, the sheer volume of collected and disseminated data provides detailed records on individuals. Such microdata is indispensable for analytical tasks in data-driven projects, assisting in research and strategy development. However, the continuous analysis and reuse of this data on a vast scale challenges the privacy of individuals, sparking international legal debates and heightening concerns about how institutions handle personal data.

In response to these concerns, a range of data privacy regulations have emerged. The GDPR\cite{GDPR2016a} was introduced to streamline data privacy laws across Europe, followed by the CCPA for Californians and the LGPD for Brazilians. These legislations compel organizations to adopt appropriate measures for processing personal data. While the data subject, the person to whom the data refers, has rights such as access and erasure, data controllers, and processors, the entities managing the data, bear obligations like ensuring confidentiality and breach notifications.

Balancing data confidentiality with its utility is a complex endeavor. High confidentiality might inhibit the extraction of valuable insights, whereas minimal privacy could lead to breaches and re-identification. This balancing act presents a dichotomy in research between data encryption and data transformation. While encryption using cryptographic methods offers robust security, it poses challenges in data manipulation and analysis. Data transformation, on the other hand, modifies data for public release while maintaining its interpretability, a crucial factor when sharing information.

This survey aims to comprehensively review the Statistical Disclosure Control issue and privacy-preserving techniques for data transformation. We focus on methods that prioritize interpretability and serve as foundational tools for future computational analysis. Among our core contributions are a detailed definition of microdata de-identification, a taxonomy of privacy-preserving techniques, metrics for information loss and predictive performance, and reviews of existing studies.

Previous work in this domain has either been segmented or outdated. This survey integrates and updates prior studies, providing a nuanced analysis of the efficacy of privacy-preservation techniques in the contemporary setting.

\section{ PRIVACY RISKS}
\label{sec:headings}

In the realm of microdata, where specific attributes define datasets, the choice of potent privacy-preserving techniques becomes vital for data controllers in achieving de-identified datasets. These transformations commonly revolve around three primary privacy threats:

Singling Out: The risk that an intruder might isolate specific records to uniquely identify a data subject.
Linkability: The danger of connecting or correlating two or more records of an individual or a set of individuals.
Inference: The potential to derive the value of one attribute based on the values of other attributes.
Moreover, the rise in attempts to breach confidential information has spotlighted other privacy threats, albeit less renowned, which we'll touch upon.

To determine the effectiveness of a chosen privacy-preserving method, data controllers use distinct privacy measures suited for each threat mentioned above. However, measuring disclosure risk poses challenges. Data breaches often result from intruders leveraging external information, information that the data controller might not have anticipated. To combat this, data controllers must hypothesize about potential external knowledge sources to estimate risk. This often involves exploring various scenarios, such as different Quality Indicators (QI) that intruders might be privy to, or considering whether intruders are aware of study participants. To err on the side of caution, the most robust stance is assuming a maximum-knowledge intruder, implying they're aware of all initial attribute values. This allows for the most accurate risk estimate.

Delving deeper, we identify four prominent disclosure risks from literature, each with its measures:
1. Identity Disclosure\cite{doyle2001confidentiality}: This occurs when an intruder matches QI values to discern a specific record in the revealed dataset is linked to a particular individual.
2. Attribute Disclosure\cite{Duncan1989TheRO}: Here, an intruder determines additional attributes about a data subject from the disclosed data.
3. Inferential Disclosure: This risk involves an intruder deducing specific private data about a subject, backed by the statistical characteristics of the released data. Importantly, inferential disclosure can even jeopardize those not included in the dataset. However, since inference techniques focus more on collective rather than individual behavior or attributes, it's usually not a focal point in Statistical Disclosure Control (SDC) for microdata. As such, we won't be elaborating on measures for inferential disclosure in this discussion.
4. Membership Disclosure: This is when an intruder can discern whether data pertaining to a specific individual is included in the dataset or not. Presently, there are only a handful of metrics, like $\delta$-presence, which is designed to limit the probability of deducing any potential data subject's record within a specific $\delta$ range. Given the limited focus on this type of risk in the current context, we'll omit its detailed measures moving forward.

\section{PRIVACY-PRESERVING TECHNIQUES}
The crux of de-identification lies in revealing data that aids organizations, administrations, and businesses in informed decision-making while safeguarding sensitive particulars of specific data subjects. The quest, therefore, revolves around unearthing methodologies that trim down disclosure risks while retaining the potency for statistical evaluations and data mining. This delicate balance between safeguarding data and maintaining its utility has fueled the quest for pioneering privacy-preserving approaches, and in some instances, revamping of existing ones.

Pioneers in the domain, Willenborg and Waal, laid down foundational principles for safeguarding microdata, introducing a taxonomy of privacy-preserving methods based on microdata traits. This taxonomy segregates techniques into:

Non-perturbative Techniques: These are defined by reducing intricacy or even omitting information altogether. The key aspect here is the absence of inconsistencies, ensuring that intruders cannot pinpoint records that have undergone transformation.

Perturbative Techniques: As the name suggests, these revolve around tweaking or altering the information. The side effect of this technique is that it can create inconsistencies in data. Such inconsistencies might pique the interest of an intruder, drawing them to speculate on records that might have been altered and thereby attempt to restore them to their original state.

While these classifications have dominated academic discussions, there's a fresh category believed to fall under Statistical Disclosure Control (SDC): De-associative Techniques. These are designed to sever the link between Quality Indicators (QI) and sensitive traits. Instead of releasing a unified table, the strategy is to release two distinct tables, one with QI and the other with sensitive attributes.

\subsection{Differential Privacy (DP)}

Differential Privacy (DP)\cite{Dwork2006,Dwork2008} aims to protect individual data privacy in data analysis. 
It adds noise to the data or responses to queries to obscure individual entries. Key Concepts:
- Epsilon ($\epsilon$): A parameter that measures the privacy guarantee. Smaller $\epsilon$ means better privacy.
- Delta ($\delta$): A small probability of privacy failure, usually close to zero.
- Random Noise: Added to responses based on the sensitivity of the query and $\epsilon$.

DP's Basic Principle:
For two datasets D and D' differing by one individual, and for any subset of outcomes S:
  $$Pr[M(D) \in S] \leq e^{\epsilon}  Pr[M(D') \in S] + \delta$$

This ensures that the presence/absence of any individual in the dataset does not significantly
affect the outcome, thus preserving privacy.

\subsection{Local Differential Privacy (LDP)}
Local Differential Privacy (LDP) is a variant of DP where privacy is guaranteed 
before any data leaves an individual's device. \cite{prolong_DP,Conti_release_LDP, 203872, Cormode:2018:PSL:3183713.3197390}
In LDP, each user perturbs their data locally before sending it to the aggregator.
This ensures privacy protection at the individuall evel.

Key Concepts:
- Local Randomization: Users apply a randomized algorithm to their data locally.
- No Trust in Data Collector: Unlike traditional DP, LDP does not require trusting the data collector
  with original data.

LDP's Basic Principle:
For any individual's data x and x', and for any subset of outcomes S:
  $$Pr[LM(x) \in S] \leq e^{\epsilon}  Pr[LM(x') \in S]$$

This ensures that each individual's data remains private, even from the data collector.

\section{State of the Art Deployment}
We delve into three tangible implementations of LDP algorithms utilized for gathering popularity metrics. These encapsulate the evolution of ideas within computer science research, as well as the advancements made thereafter.

1. RAPPOR by Google\cite{Rappor}: Merging the concepts of Randomized Response and Bloom Filters, RAPPOR efficiently compresses extensive sets. It's primarily designed to pinpoint frequently visited web destinations (URLs) while safeguarding individual browsing patterns. The same group furthered their research, illustrating an approach to discern popular websites without pre-existing knowledge of their URLs.

2. Apple's DP Mechanism\cite{apple_learning_privacy}: Unveiled in 2016, this implementation is detailed in a patent application and a later white paper. The method integrates several innovations: the deployment of the Fourier transform to disseminate signal details and the use of sketching methods to trim down the massive domain's dimensionality. Concurrently, a burgeoning literature trend has approached the task of pinpointing Differentially Private Heavy Hitters, continually honing and enhancing these methods.

3. Microsoft's Telemetry Collection: This strategy utilizes histograms coupled with set random numbers, facilitating data collection over extended periods.

\section{LDP variants and mechanisms}
In this section, we delve into the diverse variants and mechanisms of LDP designed to improve the trade-off between utility and privacy for various applications.

\subsection{Variants and Mechanisms of LDP}


\textbf{Geo-indistinguishability}
Building on the principles of differential privacy, Geo-indistinguishability was developed primarily for location privacy. It leverages the geographical distance inherent in data \cite{10.1145/2508859.2516735}.

\textbf{BLENDER Model}
Taking a step towards combining global DP and LDP, we have the BLENDER model. This model, designed to balance data utility while maintaining privacy, classifies users based on trust factors and optimally utilizes data from both categories.

\textbf{Utility-optimized LDP (ULDP)\cite{10.5555/3361338.3361468}}
While traditional LDP tends to add noise uniformly, the ULDP model recognizes that not all data elements share the same sensitivity level. This model proposes an optimization that caters to the varying sensitivity of data points.

\textbf{Input-Discriminative LDP (ID-LDP)}
For a more granular understanding of data sensitivity, ID-LDP is introduced. Unlike ULDP
, which classifies data into broad categories of sensitive and non-sensitive, ID-LDP acknowledges the varying levels of sensitivity within data. By assigning different privacy budgets based on this sensitivity, it allows for a more nuanced privacy mechanism.

\textbf{Local Information Privacy}
Progressing further, the Local Information Privacy (LIP) comes into the picture\cite{LIP1,LIP2}. Initially introduced as a version of LDP aware of prior data, it was later adapted by Jiang et al. to assume only a partial prior-aware state \cite{jiang2019local}.

\textbf{Sequential Information Privacy}
Building upon the concepts of LIP, the Sequential Information Privacy (SIP)\cite{jiang2023online} focuses on the unique challenges posed by sequential or time series data. It seeks to measure the privacy leakage for such data, drawing similarities with the principles of LDP.

\textbf{CLDP}\cite{DBLP:journals/corr/abs-1905-06361}
Addressing one of the limitations of LDP, Gursoy et al. recognized the decreased utility of LDP in scenarios with fewer users. This led to the introduction of the metric-based model, condensed local differential privacy (CLDP) \cite{DBLP:journals/corr/abs-1905-06361}.


\textbf{Parameter Blending Privacy (PBP)}
Lastly, the Parameter Blending Privacy (PBP) model offers an advanced and flexible variant of LDP. It allows for the blending of privacy parameters to achieve an optimal privacy setting \cite{DBLP:journals/corr/abs-1905-06361}.

\textbf{Membership Privacy \cite{10.1145/2508859.2516686}}
In Differential Privacy or Local Differential Privacy's definition, the likelihood ratio must be upper and lower bounded simultaneously. However, this constraints is relaxed in the definition of membership privacy, which only considers one side of the boundaries, and thus providing either positive or negative memebership privacy protection.

\noindent In conclusion, the variety of LDP mechanisms and variants showcase the ongoing efforts to balance utility and privacy in different application scenarios. Each variant brings its own set of advantages, catering to unique challenges in the realm of privacy preservation.

\section{Conclusion}
In conclusion, the rapid expansion of microdata usage has necessitated the development of robust privacy protection measures. Statistical Disclosure Control (SDC) has emerged as a critical approach to de-identify sensitive information, balancing the privacy of individuals with the need for maintaining data utility. Our comprehensive survey has provided an in-depth look at the current state of SDC techniques, evaluating their effectiveness in different scenarios, and the trade-offs they present between confidentiality and the fidelity of data analysis. Although significant progress has been made in the field, the quest for an ideal equilibrium between data privacy and utility remains ongoing. As the landscape of data privacy evolves, so must the strategies and methods for protecting individual privacy without compromising the quality of data-driven insights. Future research directions are plentiful, with many unresolved issues offering opportunities for innovation. The continuous refinement of SDC techniques and the exploration of new methodologies will be crucial in advancing the field and achieving the dual goals of robust privacy protection and retention of data interpretability.

\bibliographystyle{unsrt}  
\bibliography{references,ref}

\end{document}